\documentclass[prb,twocolumn,unsortedaddress,superscriptaddress]{revtex4-1}
\usepackage[T1]{fontenc}
\usepackage[utf8]{inputenc}
\usepackage{graphicx}
\usepackage{amsmath}
\usepackage{color}
\usepackage{comment}
\usepackage{hyperref}
\usepackage[left=1.5cm,right=1.5cm,top=2cm,bottom=2cm]{geometry}

\newcommand{\mc}[1]{\multicolumn{1}{c}{#1}}
\newcommand{\ph}[1]{\phantom{#1}}
\begin{document}

\title{Towards the Efficient Local Tailored Coupled Cluster Approximation \\
        and the Peculiar Case of Oxo-Mn(Salen)}

\author{Andrej Antal\'{i}k}
\email{andrej.antalik@jh-inst.cas.cz}
\affiliation{J. Heyrovsk\'{y} Institute of Physical Chemistry, Academy of Sciences of the Czech \mbox{Republic, v.v.i.}, Dolej\v{s}kova 3, 18223 Prague 8, Czech Republic}
\affiliation{Faculty of Mathematics and Physics, Charles University, Ke Karlovu 3, 12116, Prague 2, Czech Republic}

\author{Libor Veis}
\email{libor.veis@jh-inst.cas.cz}
\affiliation{J. Heyrovsk\'{y} Institute of Physical Chemistry, Academy of Sciences of the Czech \mbox{Republic, v.v.i.}, Dolej\v{s}kova 3, 18223 Prague 8, Czech Republic}

\author{Ji\v{r}\'{i} Brabec}
\email{jiri.brabec@jh-inst.cas.cz}
\affiliation{J. Heyrovsk\'{y} Institute of Physical Chemistry, Academy of Sciences of the Czech \mbox{Republic, v.v.i.}, Dolej\v{s}kova 3, 18223 Prague 8, Czech Republic}

\author{\"Ors Legeza}
\email{legeza.ors@wigner.mta.hu}
\affiliation{Strongly Correlated Systems ``Lend\"{u}let'' Research group, Wigner Research Centre for Physics, H-1525, Budapest, Hungary}

\author{Ji\v{r}\'{i} Pittner}
\email{jiri.pittner@jh-inst.cas.cz}
\affiliation{J. Heyrovsk\'{y} Institute of Physical Chemistry, Academy of Sciences of the Czech \mbox{Republic, v.v.i.}, Dolej\v{s}kova 3, 18223 Prague 8, Czech Republic}

\date{\today}

\begin{abstract}

  We introduce a new implementation of the coupled cluster method tailored by matrix product
  states wave functions (DMRG-TCCSD), which employs the local pair natural orbital approach (LPNO).
  By exploiting locality in the coupled cluster stage of the calculation, we were able to remove some
  of the limitations that hindered the application of the canonical version of the method to larger systems
  and/or with larger basis sets.
  We assessed the accuracy of the approximation using two systems: tetramethyleneethane (TME) and oxo-Mn(Salen).
  Using the default cut-off parameters, we were able to recover over 99.7\% and 99.8\% of canonical correlation
  energy for the triplet and singlet state of TME respectively.
  In case of oxo-Mn(Salen), we found out that the amount of retrieved canonical correlation energy depends on the
  size of the active space (CAS) -- we retrieved over 99.6\% for the larger 27 orbital CAS and over 99.8\%
  for the smaller 22 orbital CAS.
  The use of LPNO-TCCSD allowed us to perform these calculations up to quadruple-$\zeta$ basis set amounting 
  to 1178 basis functions.
  Moreover, we examined dependance of the ground state of oxo-Mn(Salen) on CAS composition.
  We found out that the inclusion of 4d$_{xy}$ orbital plays an important role in stabilizing the singlet
  state at the DMRG-CASSCF level via double-shell effect.
  However, by including dynamic correlation the ground state was found to be triplet regardless of the size
  of the basis set or composition of CAS, which is in agreement with previous findings by canonical DMRG-TCCSD
  in smaller basis.
\end{abstract}


\maketitle

\section{Introduction}
\label{sec_intro}
% CC intro
Since its introduction to quantum chemistry \cite{cizek-original}, the coupled cluster (CC) approach has become
one of the most widely used methods for the accurate calculations of dynamic correlation.
It offers numerous favorable properties, such as compact description of the wave function, size-extensivity,
invariance to rotations within occupied or virtual orbital subspaces and also a systematic hierarchy
of approximations converging towards the full configuration interaction (FCI) limit \cite{gauss-encyclop}.
For instance, the CCSD(T) method \cite{ccsdtparent1}, which includes connected single-, double- and perturbative
triple excitations, is notoriously referred to as the gold standard of quantum chemistry \cite{gauss-encyclop}.

% static corr, MR CC
Although the CC method performs well for single reference molecules, it becomes fairly inaccuarate or breaks down completely
for systems with strongly correlated electrons.
Such systems are multireference in nature since they include quasi degenerate frontier orbitals, which are common
during dissociation processes, in diradicals, or compounds containing transition metals.
Over the years, numerous efforts to generalize the CC ansatz and thus overcome this drawback
gave rise to a broad family of multireference CC methods (MRCC) \cite{bartlett-musial2007,tew10,lyakh_2012}.

% TCC
One such approach, aiming to include static correlation in the CC scheme is to employ
a different method like complete active space self-consistent field (CASSCF) or multireference configuration
interaction (MRCI) in order to extract the information about the most important excitations
\cite{paldus-externalcorr,paldus-externalcorr2,paldus-externalcorr-new1,paldus-externalcorr-new2,
ec-1,tobola1996,peris1997,kinoshita_2005,cyclobut-tailored-2011,melnichuk-2012,melnichuk-2014,
Piecuch1993,Piecuch1994,Adamowicz1998,Piecuch2010,Piecuch2002_1,Piecuch2002_2,Kowalski2002,Wloch2006,
Lodriguito2006,Piecuch2004,Kowalski2001,Kowalski2000}.
The retrieved information can be then introduced to a CC calculation as an external correction.
One of such methods is tailored CC with single and double excitations (TCCSD) proposed by
Kinoshita et al. \cite{kinoshita_2005},
which draws on the split-amplitude ansatz, in which the amplitudes corresponding to single and double
excitations are split into two parts.
The active part is treated by complete active space configuration interaction (CAS-CI)
and external amplitudes are iterated using the standard CCSD framework.
We recently extended this approach by using the density matrix renormalization group (DMRG)
method to obtain the active space amplitudes \cite{veis_2016}.

% DMRG + dynamic corr
The DMRG method, which originated in solid-state physics \cite{White-1992a,White-1992b,White-1993}, is nowadays well established
in quantum chemistry for the treatment of strongly correlated systems
\cite{white_1999,Chan-2002a,Legeza-2003a,Legeza-2008,Marti-2010c,Chan-2011,Wouters-2014e,legeza_review,Yanai2014}.
As a numerical approximation to full configuration interaction (FCI), it can handle significantly
larger active spaces compared to the conventional method.
However, even then the prohibitive scaling does not allow to include dynamic correlation
and it is therefore necessary to employ some "post-DMRG" procedure.
Many different attemps has been made to tackle this limitation 
for example
DMRG-CASPT2\cite{kurashige_2011},
Cholesky decomposition DMRG-NEVPT2\cite{Freitag2017},
DMRG-icMRCI\cite{Saitow-2013},
canonical transformation\cite{yanai_2010},
matrix product state (MPS) based formulation of multireference perturbation theory\cite{Sharma-2014b},
DMRG pair-density functional theory\cite{sharma_2019},
and also our aforementioned CC tailored by MPS wave functions (DMRG-TCCSD)\cite{veis_2016}.

% Local approaches + our choice
Even though the DMRG-TCCSD method offers a reasonably efficient treatment of both static and dynamic correlation\cite{veis_2018},
its applications to larger systems is hampered by the infavorable scaling of the CCSD part of the calculation.
To remove this restriction, we decided to implement the method using a local approach.
Out of many possibilities how to exploit locality
\cite{Flocke2004,Fedorov2005,Kobayashi2008,Scuseria1999,Li2001,Stoll1992,Kristensen2011,Li2009},
we opted for pair natural orbitals (PNO) based methods,
in particular, the local pair natural orbital (LPNO) approach\cite{local_cc_neese,local_crcc,hansen_2011}.

% PNO
The PNOs were introduced by Edmiston and Krause and were shown to provide a compact parametrization of the virtual space\cite{Edmiston1965}.
Over the years, several correlation methods that made use of advantages the approach offered were developed
\cite{Meyer1971,Meyer1974,Taylor1981,Fink1993}, but the full potential of PNOs could have been unleashed
only due to more recent advances in modern hardware and modern integral transformation technologies,
particularly the density fitting or resolution of the identity methodology\cite{Vahtras1993}.
The approach was further extended to domain LPNO (DLPNO), with even more favourable scaling \cite{Riplinger2013}.
Apart from single-reference methods, the LPNO and DLPNO methodologies were also succesfully applied
to multireference CC techniques \cite{Demel2015,Lang2017,Brabec2018,Lang2019}

%Step towards DLPNO-TCCSD
In this paper, we contribute to these efforts by implementing the LPNO version of DMRG-TCCSD.
We demonstrate the properties of the method on two systems which were previously studied with its canonical
implementation\cite{veis_2016,veis_2018}.
First, we used tetramethyleneethane (TME) as a benchmark system to estimate the amount of correlation energy
possible to retrieve by LPNO approach compared to the canonical version of the method.
Then, we performed a similar test on oxo-Mn(Salen) in a double-$\zeta$ basis set and subsequently
performed calculation using up to a quadruple-$\zeta$ basis set, far beyond the capabilities of the canonical
implementation.
This way, we were able to investigate the effect of dynamic correlation in the basis sets of size
previously unfeasible.
Moreover, we explored the impact of active space composition on the spin state ordering of oxo-Mn(Salen)
in order to shed light into previously varying claims about its ground state.

In the rest of this paper, we will use the acronym TCCSD($e$,$o$) to denote a DMRG-TCCSD calculation, in which the active space
of DMRG consists of $e$ electrons in $o$ orbitals. In the same manner, LPNO-TCCSD($e$,$o$) denotes the calculation with
the CC part perfomed employing the LPNO approach.

\section{Theory and Implementation}
\label{sec_theory}

\subsection{DMRG-based Tailored Coupled Clusters}

The tailored coupled cluster method, which belongs to the class of externally corrected methods, employs
the split-amplitude wave function ansazt proposed by Kinoshita et al. \cite{kinoshita_2005}
\begin{equation}
  | \Psi_\text{TCC} \rangle = e^T | \Phi_0 \rangle
                            = e^{T_\text{ext}+T_\text{CAS}} | \Phi_0 \rangle
                            = e^{T_\text{ext}}e^{T_\text{CAS}} | \Phi_0 \rangle,
\end{equation}
where the cluster operator $T$ is split into two parts:
$T_\text{CAS}$ which contains the active amplitudes obtained from an external calculation and
$T_\text{ext}$ which contains the external amplitudes, with
$| \Phi_0 \rangle$ being the reference wave function.

In our implementation, we employed the DMRG method to obtain the active amplitudes.
Using the DMRG algorithm we first optimize the wave function, which is provided in
matrix product state (MPS) form
\begin{equation}
  | \Psi_\text{MPS} \rangle = \sum_{\{\alpha\}}
  \mathbf{A}^{\alpha_1} \mathbf{A}^{\alpha_2} \cdots \mathbf{A}^{\alpha_k}
  | \alpha_1 \alpha_2 \cdots \alpha_k \rangle,
\end{equation}
where $\alpha \in \{ |-\rangle, |\downarrow\rangle, |\uparrow\rangle, |\downarrow\uparrow\rangle \}$
and $\mathbf{A}^{\alpha_i}$ are MPS matrices.
These are then contracted to obtain CI coefficients for single and double excitations $C$
\cite{moritz_2007,boguslawski_2011}.
Using the relations between CI and CC coefficients
\begin{align}
  T^{(1)}_\text{CAS} &= C^{(1)}, \\
  T^{(2)}_\text{CAS} &= C^{(2)} - \frac{1}{2}[C^{(1)}]^2,
\end{align}
we are able to acquire their rescpective amplitudes, which are subsequently introduced into
a CC calculation.
At this point, these active amplitudes are kept frozen, while the remaining amplitudes $T_\text{ext}$
are optimized by solving the equations
\begin{align}
  \langle \Phi_i^a | He^{T_\text{ext}}e^{T_\text{CAS}} | \Phi_0 \rangle_c &= 0 \quad
  \{ i,a \} \not\subset \text{CAS} \label{eq_TCC_singles} \\
  \langle \Phi_{ij}^{ab} | He^{T_\text{ext}}e^{T_\text{CAS}} | \Phi_0 \rangle_c &= 0 \quad
  \{ i,j,a,b \} \not\subset \text{CAS} \label{eq_TCC_doubles}
\end{align}
analogous to the standard CCSD equations.
This way, the active amplitudes account for static correlation and by optimizing the external amplitudes, we are able
to recover the remaining dynamic correlation.

\subsection{The LPNO Approach for DMRG-TCCSD}
The first step in LPNO methods is the localization of internal orbitals and subsequent transformation
of virtual space to a PNO basis.
This process can be divided into three distinct steps.
First, based on the MP2 calculation, pair energies and MP2 amplitudes are calculated.
From these, pair density matrices are constructed for pairs of occupied orbitals and by their subsequent
diagonalization PNOs are acquired.

Since approximations are made during this procedure, the accuracy of the method is controlled
mainly by two cut-off parameters.
The first, $T_\text{CutPairs}$, limits the number of occupied orbital pairs chosen for CCSD correlation
treatment based on their respective pair energies.
The remaining weak pairs are then treated solely at the MP2 level.
The second, $T_\text{CutPNOs}$, determines the truncation of the PNO expansion for a given pair, based on the PNO
occupation numbers.

In order to make the LPNO-TCCSD method function properly, we considered crucial to maintain the properties of
the canonical TCCSD method as well as the behavior of the original LPNO-CCSD method. 
Just like the latter, it should maintain smooth dependance of retreived canonical correlation energy with
respect to a change in the cut-off parameters \cite{}.
Moreover, it is necessary to perform the PNO transformation in such a way that the active orbitals in the new basis
exactly align with the orbitals in the original MO basis.
Failing to do so would result in a mismatch between the orbitals and imported amplitudes.
For the same reason, it is required for active orbitals to pass both $T_\text{CutPairs}$ and
$T_\text{CutPNOs}$ screenings.

Because we implemented LPNO-TCCSD in the spin-unrestricted version of the LPNO-CCSD code, the following
steps will be described in the respective formalism \cite{hansen_2011}.
Also, for more compact notation we will write 
$^{\sigma\sigma'}\!t_{ij}^{ab} = t_{i_\sigma j_{\sigma'}}^{a_\sigma b_{\sigma'}}$.

First, for each pair of occupied orbitals $ij$, we construct the MP2 amplitudes
\begin{align}
  ^{\alpha\alpha}t_{ij}^{ab} = -\frac{K_{ij}^{ab} - K_{ij}^{ba}}{f_{aa}+f_{bb}-f_{ii}-f_{jj}}, \\[1em]
  ^{\alpha\beta}t_{ij}^{ab} = -\frac{K_{ij}^{ab}}{f_{aa}+f_{bb}-f_{ii}-f_{jj}},
\end{align}
from exchange integrals $K_{ij}^{ab}$ and orbital energies $f_{pp}$.
The $\beta\beta$ case is constructed analogously as the $\alpha\alpha$ case.
Subsequently, we replace the active amplitudes, for which $\{ i,j,a,b \} \subset \text{CAS}$, 
with the amplitudes imported from a DMRG calculation.
At this point, the first cut-off parameter comes into play.
For each pair, we calculate an MP2 pair energy and if it is larger than $T_\text{CutPairs}$,
the pair is kept for further correlation treatment.
To ensure that none of the active amplitudes is discarded, we circumvent this screening for all active pairs
and keep them automatically.

In the next step, the PNOs are generated. First, for given pair $ij$
a pair density matrix is built from the matrix $\mathbf{T}^{ij}_{\sigma\sigma'}$ containing the amplitudes
$^{\sigma\sigma'}\!t_{ij}^{ab}$
\begin{align}
  \mathbf{D}^{ij}_{\alpha\alpha} &= \frac{4 (\mathbf{T}^{ij}_{\alpha\alpha})^\dagger\mathbf{T}^{ij}_{\alpha\alpha}}
  {1+2\text{Tr}((\mathbf{T}^{ij}_{\alpha\alpha})^\dagger\mathbf{T}^{ij}_{\alpha\alpha})}, \\[1em]
  ^{(\alpha)}\mathbf{D}^{ij}_{\alpha\beta} &= \frac{2 \mathbf{T}^{ij}_{\alpha\beta}(\mathbf{T}^{ij}_{\alpha\beta})^\dagger}
  {1+\text{Tr}(\mathbf{T}^{ij}_{\alpha\beta}(\mathbf{T}^{ij}_{\alpha\beta})^\dagger)}, \\[1em]
  ^{(\beta)}\mathbf{D}^{ij}_{\alpha\beta} &= \frac{2 (\mathbf{T}^{ij}_{\alpha\beta})^\dagger\mathbf{T}^{ij}_{\alpha\beta}}
  {1+\text{Tr}((\mathbf{T}^{ij}_{\alpha\beta})^\dagger\mathbf{T}^{ij}_{\alpha\beta})}.
\end{align}
Since two sets of PNOs are needed for $\alpha\beta$ pairs, we use the superscripts $(\alpha)$ and $(\beta)$
to distinguish between them.
Once again, the $\beta\beta$ case is analogous to the $\alpha\alpha$ case.

For an inactive pair $\{i,j\}\not\subset \text{CAS}$, we proceed directly to diagonalization of the pair matrix to solve
\begin{equation}
  \mathbf{D}^{ij} \mathbf{d}^{ij}_{\bar a} = n^{ij}_{\bar a} \mathbf{d}^{ij}_{\bar a}, \label{eq_pdm_eig}
\end{equation}
and obtain set of PNOs $\mathbf{d}^{ij}_{\bar a}$ and their respective occupation numbers
$n^{ij}_{\bar a}$, where barred index refers to the PNO basis.
This PNO expansion is then truncated based on the second cut-off parameter.
Only PNOs with occupation numbers larger than $T_\text{CutPNOs}$ are kept and the remaining
orbitals are discarded.

\begin{figure}[!t]
  \begin{center}
    \includegraphics[width=0.3\textwidth]{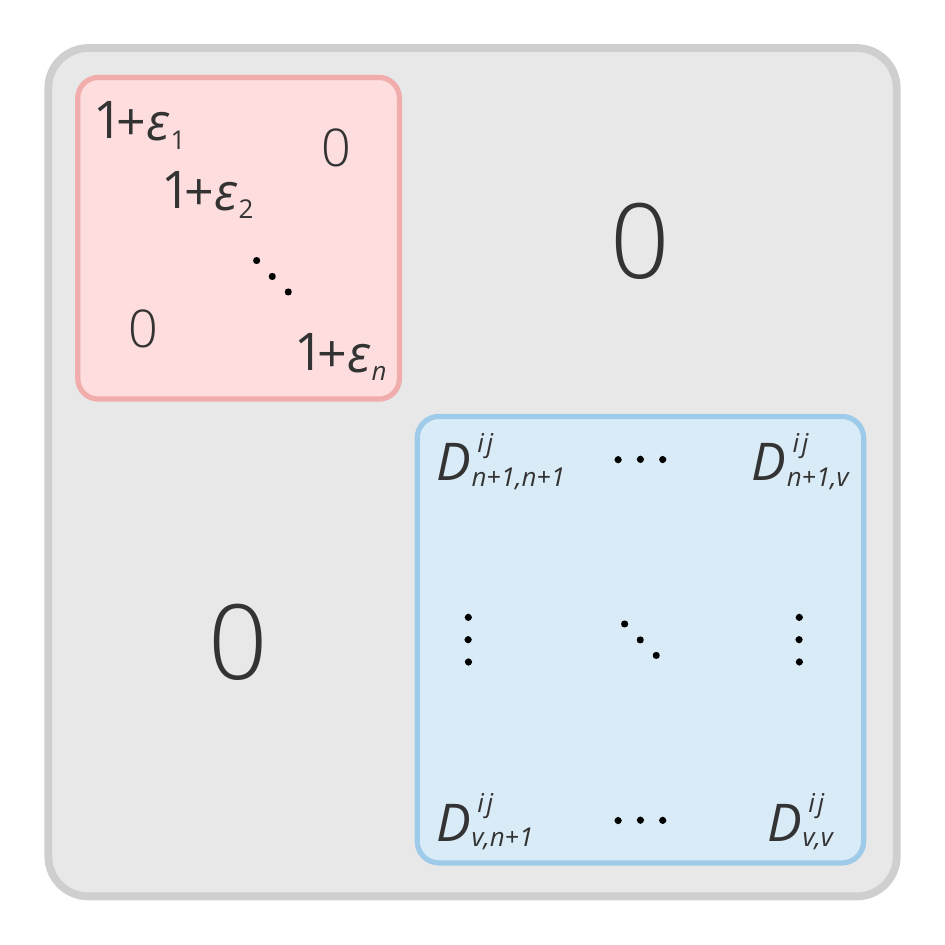}
    \caption{A pair density matrix $\mathbf{\widetilde D}^{ij}$ for an active pair $ij$.
    The number of active virtual orbitals is denoted by $n$,
    total number of virtuals by $v$. }
    \label{fig_pdm}
  \end{center}
\end{figure}
However, the process gets slightly more complicated for active pairs $\{i,j\}\subset \text{CAS}$.
To maintain the alignment between the original active orbitals in MO basis and
the active orbitals in PNO basis, it is necessary to keep their coefficients untouched during
the PNO transformation.
We achieve this by setting the active-external elements of a pair density matrix
to zero and replacing the active-active part with an identity matrix.
In order to also preserve the correct order of the active orbitals, we add "infinitesimaly" small positive
numbers $\varepsilon_a$ to the active diagonal, for which holds that $\varepsilon_a > \varepsilon_{a+1}$.
Thus, the resulting matrix has the block form (see Figure \ref{fig_pdm})
\begin{equation}
  \mathbf{\widetilde D}^{ij}  = \mathbf{D}^{ij}_\text{CAS} \oplus \mathbf{D}^{ij}_\text{ext},
\end{equation}
where
\begin{align}
  \mathbf{D}^{ij}_\text{CAS} &= \text{diag}(1\!+\!\varepsilon_1 , \,\dots , 1\!+\!\varepsilon_n), \\
  (D^{ij}_\text{ext})_{ab} &= D^{ij}_{a+n,b+n}.
\end{align}
This way, we make sure that after solving (\ref{eq_pdm_eig}) all active orbitals have the largest
eigenvalues, which are in given order at the beginning, and therefore pass the $T_\text{CutPNOs}$ screening.

The resulting equation for singly excited amplitudes remains formally the same as the equation for
the canonical method (\ref{eq_TCC_singles}).
On the other hand, the equation for doubly excited amplitudes (\ref{eq_TCC_doubles}) now becomes
\begin{equation}
  \langle \Phi_{ij}^{\bar a \bar b} | He^{T^{(1)}_\text{ext} + \bar T^{(2)}_\text{ext}}e^{T_\text{CAS}}
  | \Phi_0 \rangle_c = 0 \quad \{ i,j,a,b \} \not\subset \text{CAS},
\end{equation}
with the active amplitudes formally in PNO basis.

\section{Computational Details}
\label{sec_compdet}

The DMRG calculations were performed by Budapest QC-DMRG code \cite{budapest_qcdmrg}.
The LPNO-TCCSD method was implemented in ORCA program package \cite{orca}, which was also used to prepare the orbitals.

%ORBITALS

In case of TME, we used CASPT2(6,6)/cc-pVTZ geometries for seven values of the dihedral angle from our previous work \cite{veis_2018}.
The orbitals were prepared by CASSCF(6,6) calculation with the active space containing six 2p$_z$ orbitals on carbon atoms.

In case of oxo-Mn(Salen), we used the singlet CASSCF(10,10)/6-31G* optimized geometry by Ivanic et al. \cite{ivanic_2004}.
The orbitals were optimized using the DMRG-CASSCF method \cite{Ghosh-2008,Zgid-2008c,Yanai-2009} in Dunning's cc-pVXZ
$\text{X}\in$\{D,T,Q\} basis sets \cite{dunning_89,dunning_93,balabanov_05}.
The optimization was carried out with fixed bond dimension $M=1024$ for the smaller CAS(28,22) and $M=2048$ for CAS(28,27).
The composition of these active spaces is discussed further in the Results section.
The orbitals were then  split-localized using the Pipek-Mezey algorithm \cite{localization_pipek} in the following
orbital subspaces: internal, active doubly occupied, active singly occupied and active virtual.

%DMRG
The orbitals for DMRG were ordered using the Fiedler method \cite{barcza_2011, fertitta_2014}
combined with some manual adjustments.
All DMRG runs were initialized by CI-DEAS procedure \cite{legeza_2003b,legeza_review}.
We employed the dynamical block state selection (DBSS) procedure \cite{legeza_2003a, legeza_2004} to control the accuracy
of the larger oxo-Mn(Salen) calculations with the truncation error criterion set to $10^{-6}$.
This resulted in block dimension varying between 1000 up to 2500 block states for CAS(28,22) and up to 8200 in case of CAS(28,27).
The convergence threshold was set to energy difference between two subsequent sweeps smaller than $10^{-6}$ a.u.

%LPNO
The core electrons were kept frozen throughout all coupled cluster calculations.
Auxiliary basis sets cc-pVQZ/C and cc-pV6Z/C were used for the resolution of the identity approximation
for oxo-Mn(Salen) and TME respectively \cite{weigend_2002,bross_2013}.
The default LPNO cut-off parameters were
set to $T_\text{CutPNO}=3.33\cdot10^{-7}$, $T_\text{CutPairs}=10^{-4}$ and $T_\text{CutMKN}=10^{-3}$
and these were used unless otherwise stated.
The production runs of oxo-Mn(Salen) were performed with ORCA's TightPNO settings i.e. the cut-off parameters set to
$T_\text{CutPNO}=10^{-7}$, $T_\text{CutPairs}=10^{-5}$ and $T_\text{CutMKN}=10^{-4}$.
For calculations which purpose was to estimate the dependance of LPNO-TCCSD energies on these parameters, one
parameter was varied with remaining parameters fixed to the default value.
We assess the amount of retrieved correlation energy by LPNO approach with reference to a DMRG-TCCSD energy
calculated with the traditional TCCSD implementation.

\section{Results and Discussion}
\label{sec_results}

\subsection{Tetramethyleneethane}
\begin{figure}[!t]
  \begin{center}
    \includegraphics[width=0.25\textwidth]{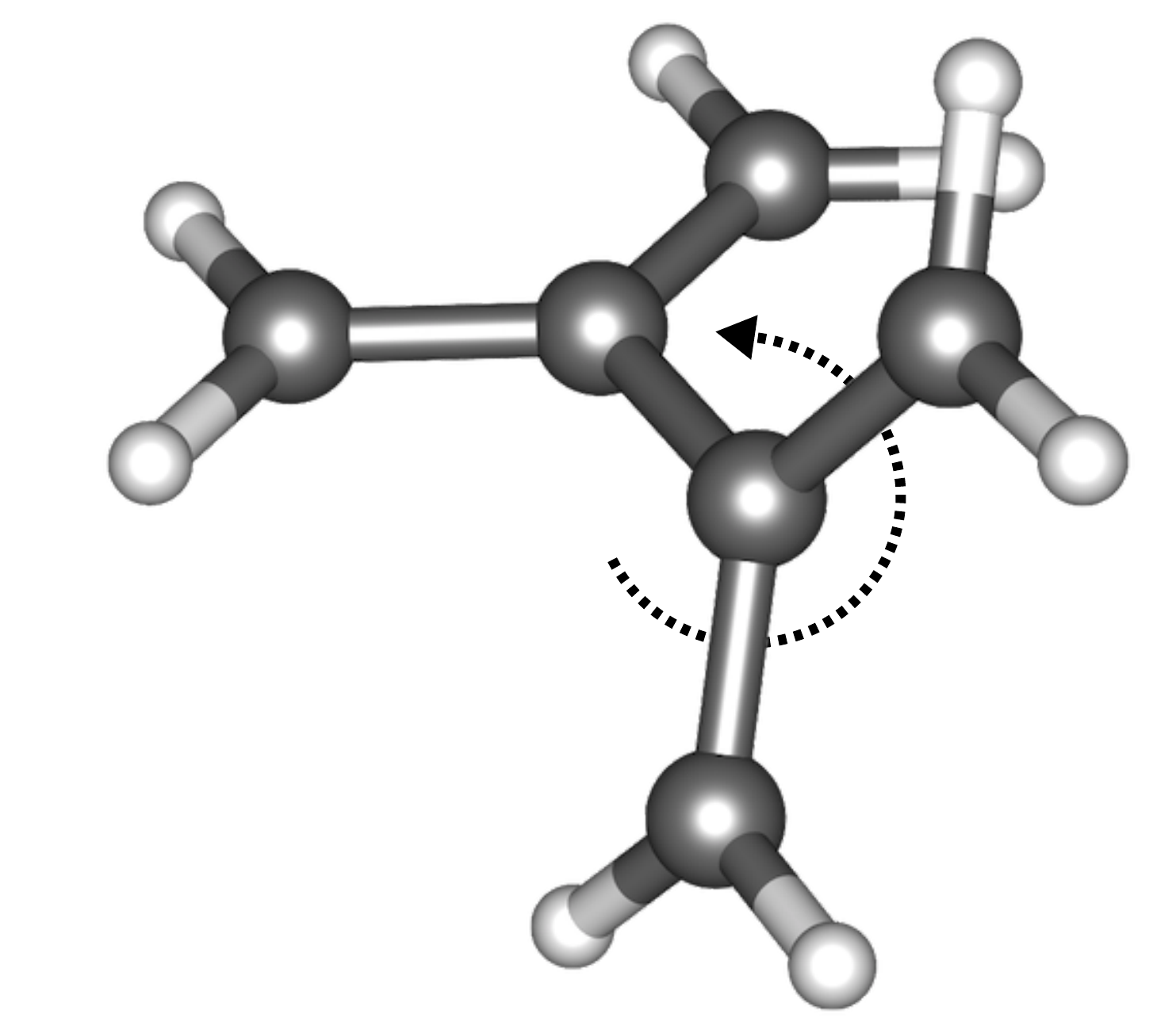}
    \caption{Dihedral rotation of tetramethylenethane.}
    \label{fig_tme}
  \end{center}
\end{figure}

Although small, the tetramethyleneethane molecule is a challenging system due to its complex electronic structure.
To correctly describe the character of its singlet state, one needs to employ a theory with a balanced description of
both static and dynamic correlation combined with a reasonably large basis set.
This is the reason, why it often serves as a benchmark system for multireference
methods \cite{pittner_2001, bhaskaran-nair_2011, chattopadhyay_2011, pozun_2013, demel_2015, veis_2018}.
Moreover, it was already a subject of our previous study with the canonical DMRG-TCCSD method \cite{veis_2018},
so it only seems natural to use this system to test the performance of the LPNO approach to TCCSD. 
For this purpose, we investigate the behavior of the approximation with respect to different geometries
corresponding to the rotation about its central C--C bond (see Figure \ref{fig_tme}) and different values of the
cut-off parameters.

We only present results for a small active space corresponding to six electrons in six 2p orbitals. This decision followed
an effort to perform the performance evaluation on three active spaces of different sizes. However, because of
the small localization
subspaces stemming from a small number of occupied orbitals, many orbitals remained rather delocalized.
This resulted in large numbers of PNOs necessary to maintain the accuracy, even for looser cut-off parameters,
which ultimately rendered the LPNO approximation useless due to its low efficiency.
Therefore, we compare the performance of LPNO-TCCSD with different sized active spaces on oxo-Mn(Salen),
which is better suited for this purpose.

At this point, we investigate the amount of correlation energy retrieved by the LPNO-TCCSD method compared to
the canonical TCCSD method.

\begin{figure*}[!ht]
  \begin{center}
    \includegraphics[width=\textwidth]{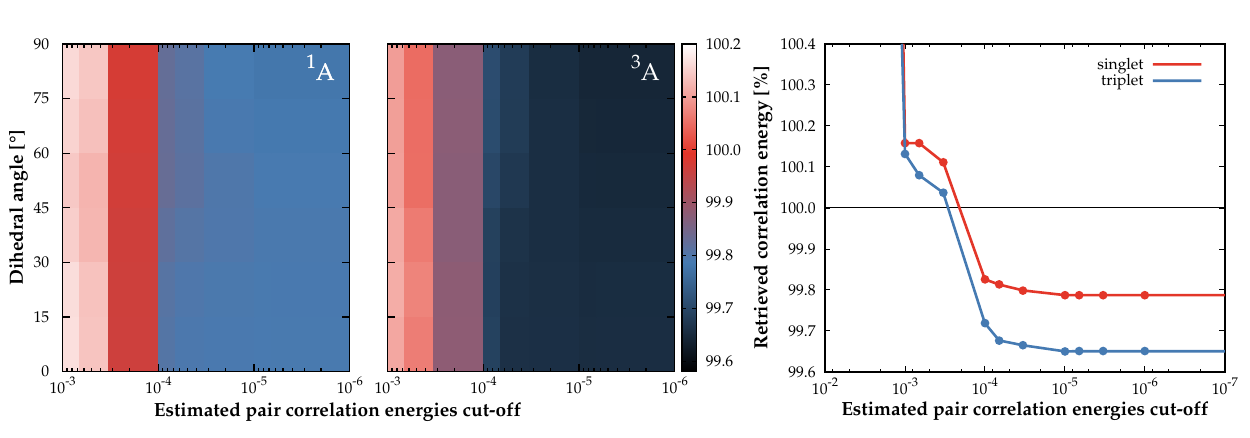}
    \caption{The percentage of correlation energy of the canonical TCCSD(6,6) calculation recovered by
    LPNO-TCCSD(6,6) in cc-pVTZ basis with respect to cut-off for estimated pair correlation energies
    $T_\mathrm{CutPairs}$. The color maps on the left show the results for seven studied geometries
    and $^1$A and $^3$A states, the plot on the right shows the results averaged over the geometries.}
    \label{fig_tme_PairsMap}
  \end{center}
\end{figure*}
\begin{figure*}[!ht]
  \begin{center}
    \includegraphics[width=\textwidth]{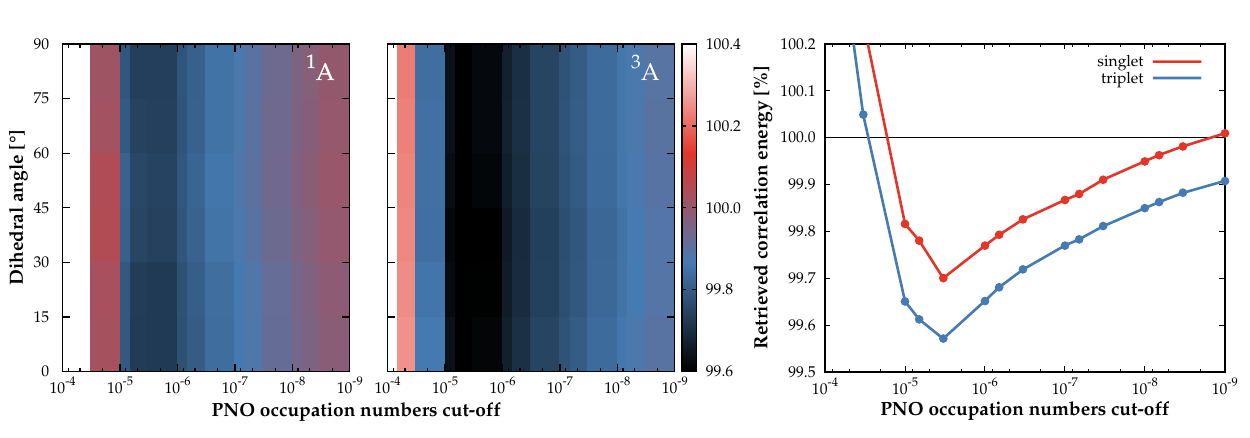} \\
    \caption{The percentage of correlation energy of the canonical TCCSD(6,6) calculation recovered by
    LPNO-TCCSD(6,6) in cc-pVTZ basis with respect to cut-off for PNO occupation numbers $T_\mathrm{CutPNO}$.
    The color maps on the left show the results for seven studied geometries and $^1$A and $^3$A states,
    the plot on the right shows the results averaged over the geometries.}
    \label{fig_tme_PNOsMap}
  \end{center}
\end{figure*}
Figure \ref{fig_tme_PairsMap} shows the dependence of the retrieved correlation energy with respect to the cut-off
parameter $T_\mathrm{CutPairs}$, which controls the number of pairs treated by CCSD
based on their estimated pair correlation energies.
Both singlet and triplet state show the desired behaviour and converge towards a certain value. For settings looser
than the default value of $10^{-4}$, the increasing number of pairs treated by MP2 results in overestimation
of the correlation energy.
As can be seen, for the triplet state the method recovers consistently about 0.1\% of correlation energy
less than for the singlet state.
This means that the energy difference between the two states is therefore off by about 0.7 kcal/mol compared
to a canonical calculation.
Regarding the consistency of the approximation across different geometries, the accuracy is mostly consistent,
but slight discrepancies are visible around the default value of the cut-off parameter.
However, the largest difference in energies at this value is less than 0.03\% canonical correlation energies,
which corresponds to an energy difference smaller than 0.2 kcal/mol.
Note that these differences are relevant for singlet or triplet calculation alone.

Figure \ref{fig_tme_PNOsMap} shows the dependence on the second studied cut-off parameter
$T_\mathrm{CutPNOs}$.
From $10^{-4}$, the recovered energy grows gradually when tightening the parameter resulting in 99.83\% and 99.72\%
of correlation energy recovered at the default value for the singlet and triplet states respectively. 
Again the difference between the geometries is less than 0.03\% for each state alone.
For the smallest values the singlet calculation retrieves more than 100\% of correlation energy, which
is caused by using the default setting for $T_\mathrm{CutPairs}$.
This means that more pairs are treated only by perturbative correction which overshoots the contributions
to correlation energy.

The observed difference in accuracy for these two cases then might arise from the neglect of some of the terms in the
current LPNO implementation.
However, this behavior should be eradicated in the future DLPNO implementation of the method.

\subsection{oxo-Mn(Salen)}
The oxo-Mn(Salen) molecule has been a subject of numerous computational studies motivated mainly by its role
in catalysis of the enantioselective epoxidation of unfunctional olefins \cite{jacobsen_1990,katsuki_1990}.
Moreover, its closely lying singlet and triplet states pose a considerable challenge for multireference methods.
Over the years, several multireference studies has been published \cite{ivanic_2004,sears_2006,ma_2011}, some
of which employed the DMRG method \cite{wouters_2014,olivares-amaya_2015,stein_2016} and recently the first
DMRG results with dynamic correlation treatment were presented \cite{veis_2016,sharma_2017}.
Our aim was to contribute to these efforts by exploring
the effect of the active space and basis set dependence.
on the character of the ground state.
With our LPNO implementation we were able to study the effect of dynamic correlation up to the quadruple-$\zeta$ basis set.
This would not be possible without the LPNO approximation, since the cc-pVQZ basis for this systems amounts to
1178 basis functions.

\begin{figure}[!t]
  \begin{center}
    \includegraphics[width=0.45\textwidth]{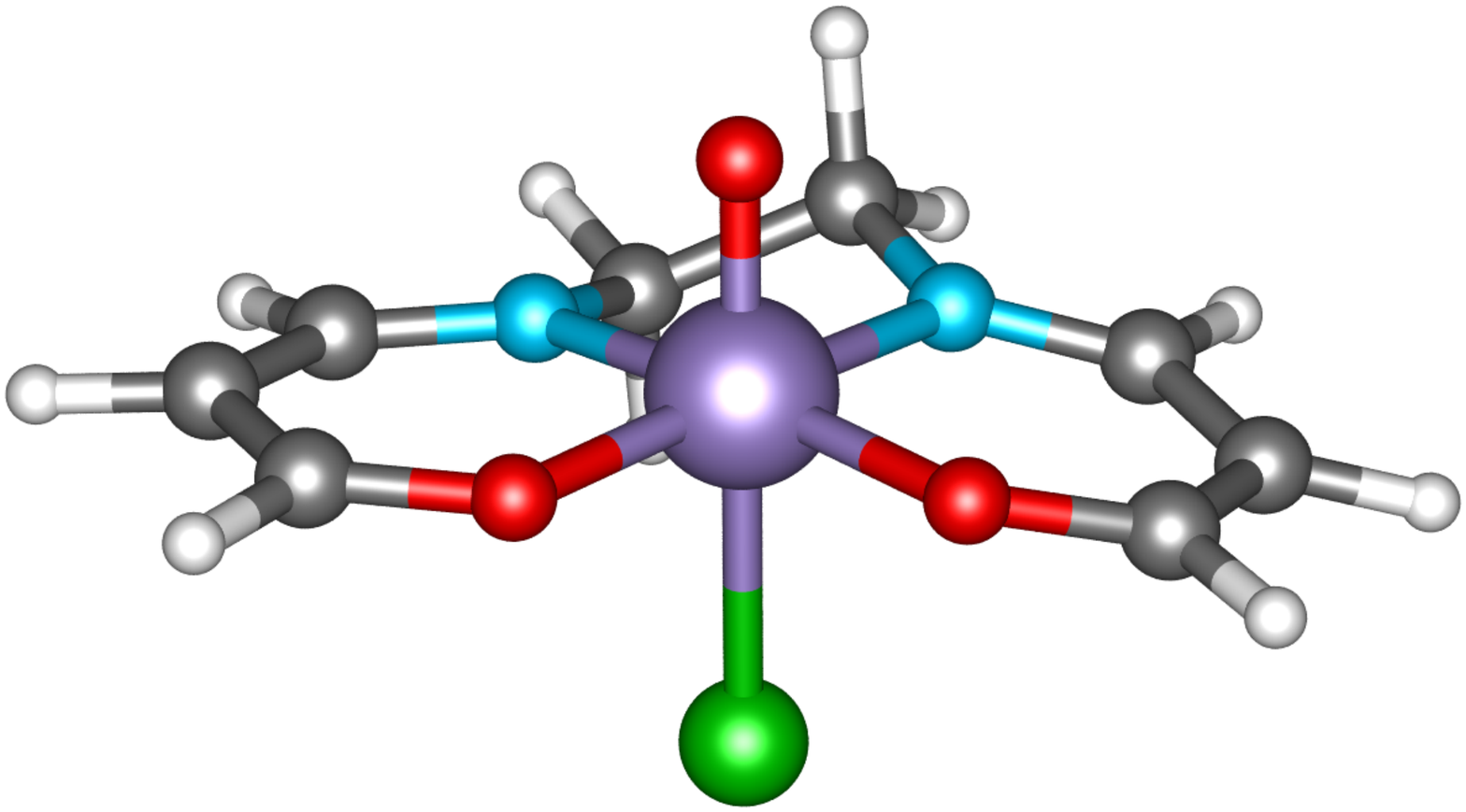} 
    \caption{A molecule of oxo-Mn(Salen)}
    \label{fig_oxo}
  \end{center}
\end{figure}

% ACTIVE SPACE selection
In order to assess the accuracy of the LPNO-TCCSD method with respect to active spaces of different size
and investigate the different ground states reported at the CASSCF level, we selected two active spaces.
In accordance with the study by Wouters et al. \cite{wouters_2014},
the smaller CAS(28,22) consists of
ten $\pi$ orbitals on equatorial conjugated rings (C, N and O atoms),
five 3d orbitals on Mn,
three 2p orbitals on the axial O atom and
four 2p orbitals on equatorial N and O atoms forming $\sigma$ bonds with the Mn atom.
On top of these, we added extra five orbitals on Mn resulting in CAS(28,27), namely
4d$_{xy}$, 4d$_{yz}$, and 4p$_{x}$, 4p$_{y}$ and 4s,
which form $\sigma^*$ bonds with Mn.
The effect of inclusion of these particular orbitals is discussed further in the text.
On top of that, we also tried to add 3p orbitals on Cl to the active space, since these were included
in some of the studies \cite{olivares-amaya_2015,veis_2016} but based on the results of entanglement analysis
(one-orbital entropies) we concluded that their effect was negligible.

\begin{figure*}[!ht]
  \begin{center}
    \includegraphics[width=0.48\textwidth]{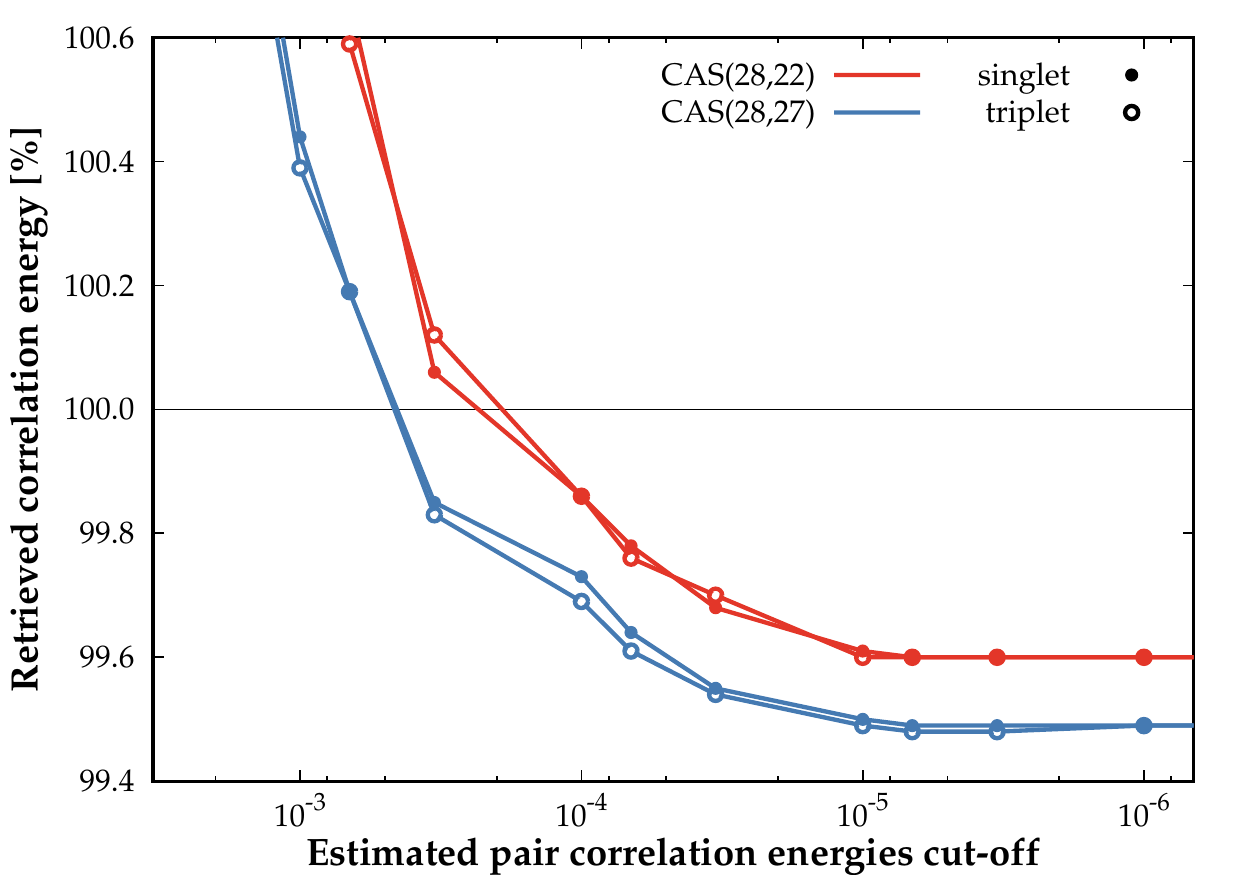}
    \includegraphics[width=0.48\textwidth]{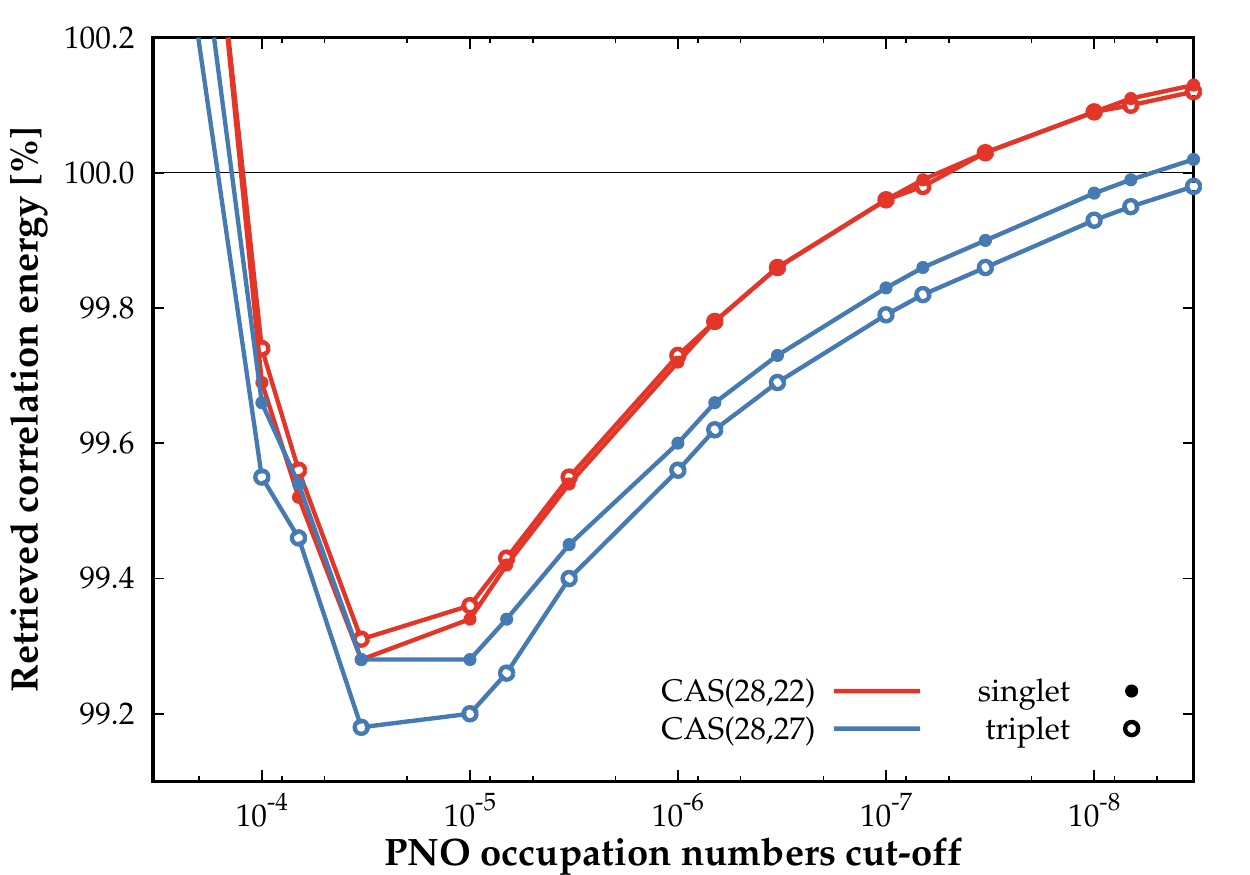}
    \caption{The percentage of correlation energy of oxo-Mn(Salen) retrieved by LPNO-TCCSD in cc-pVDZ basis
    with respect to canonical TCCSD calculations as a function of cut-off for estimated pair correlation
    energies $T_\mathrm{CutPairs}$ (left) and cut-off for PNO occupation numbers $T_\mathrm{CutPNO}$ (right).}
    \label{fig_oxo_LPNO}
  \end{center}
\end{figure*}

% LPNO performance
First, we investigate the behavior of the LPNO approximation in the smallest cc-pVDZ basis.
The dependence of recovered correlation energy for both spin states and active spaces with respect
to LPNO cut-off parameters is shown in graphs in Figure \ref{fig_oxo_LPNO}.
When varying $T_\mathrm{CutPairs}$, the curves for all four calculations converge smoothly.
The singlet and triplet curves exhibit an excellent behavior, with the errors stemming from the approximation
canceling out perfectly for both states. In case of the smaller CAS(28,22), this is valid even for less
conservative values.

The amount of correlation energy with respect to $T_\mathrm{CutPNOs}$ parameter changes smoothly towards 100\%
with smaller values.
With the default settings, we were able to recover over 99.85\% correlation energy for the smaller and over 99.78\% for
the larger active space.
The method overshoots for $T_\mathrm{CutPNOs}$ over $3.33\cdot10^{-8}$ for the same reason as with TME
i.e. the fixed $T_\mathrm{CutPairs}$ parameter.
Also the small consistent gap between singlet and triplet curves disappears with smaller
cut-off for estimated pair correlation energies.
As can be seen from Table \ref{tab_oxo_diffs}, although the absolute energies acquired with tighter cut-offs seem
to be slightly worse, these settings noticeably improve the reproduction of the canonical singlet-triplet gap.

For this system, the treatment of different spin states is balanced due to more effective cancellation of errors
compared to TME.
However, a difference in accuracy arises between the larger and smaller CAS, with slightly better retrieval
of correlation energy in case of the latter.
Similarly to TME, this can be attributed to a neglect of certain terms in the LPNO implementation of the CC method.

\begin{table}[!ht]
  \caption{Energy differences in kcal/mol between LPNO-TCCSD calculations with different settings
           of cut-off parameters and equivalent canonical calculations on oxo-Mn(Salen) in cc-pVDZ.}
  \def\arraystretch{1.2}
  \setlength{\tabcolsep}{0.5em}
  \begin{tabular}{ccccc}
                          & \multicolumn{2}{c}{CAS(28,22)} & \multicolumn{2}{c}{CAS(28,27)}  \\
                          & default           & TightPNO      & default   & TightPNO         \\ \hline
  $^1\text{A}$            & $\phantom{-}2.53$ & $4.50$        & $4.91$    & $6.47$           \\
  $^3\text{A}$            & $\phantom{-}2.47$ & $4.55$        & $5.51$    & $6.50$           \\
  $\Delta E_\text{T-S}$   & $-0.07$           & $0.04$        & $0.60$    & $0.03$   
  \end{tabular}
  \label{tab_oxo_diffs}
\end{table}

% ENERGY DIFFERENCE CAS(28,22) v. CAS(28,27)
After the accuracy assessment, we approached the actual system.
The first step was to obtain the energies of Wouters et al. \cite{wouters_2014} with our smaller CAS(28,22)
considering that this study reports on the composition of the active space in detail.
We successfully reproduced their results establishing the triplet ground state, but since several later
studies reported significantly lower energies of both states \cite{stein_2016, sharma_2017} with the CAS
of the same size or smaller, we wanted to investigate this further.

Therefore we included, on top of three orbitals forming the $\sigma^*$ bonds,
additional 4d$_{xy}$, 4d$_{yz}$ to examine how much the double-shell effect
influences the stability of these states.
With this active space, we obtained significantly lower energies (partially due to the larger number of
variational parameters) and more importantly, the two states switched resulting in the singlet ground state.
The energies can be found in Table \ref{tab_oxo}.

\begin{figure*}[!ht]
  \begin{center}
    \includegraphics[width=0.49\textwidth]{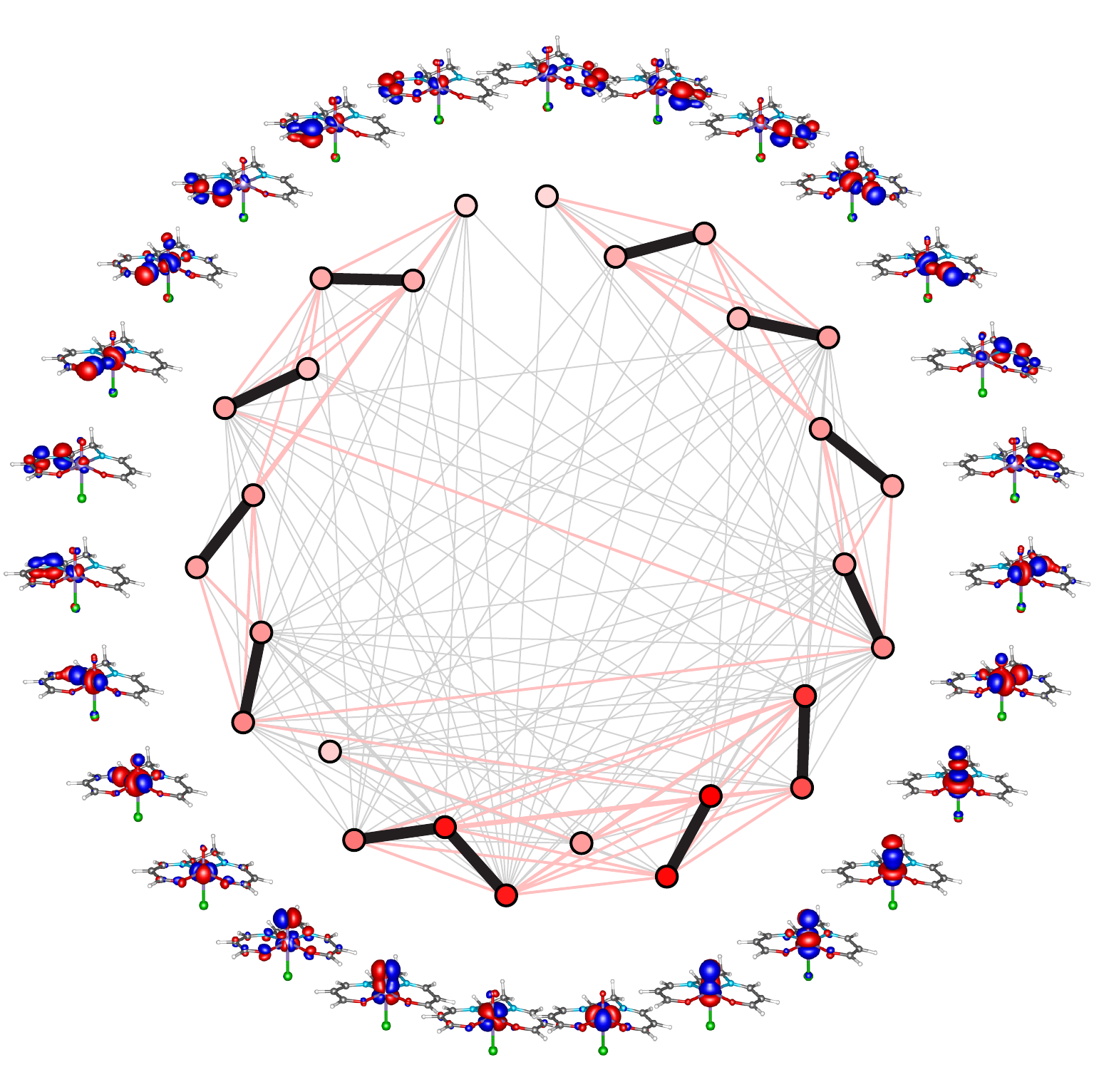}
%    \hspace{0.01\textwidth}
    \includegraphics[width=0.49\textwidth]{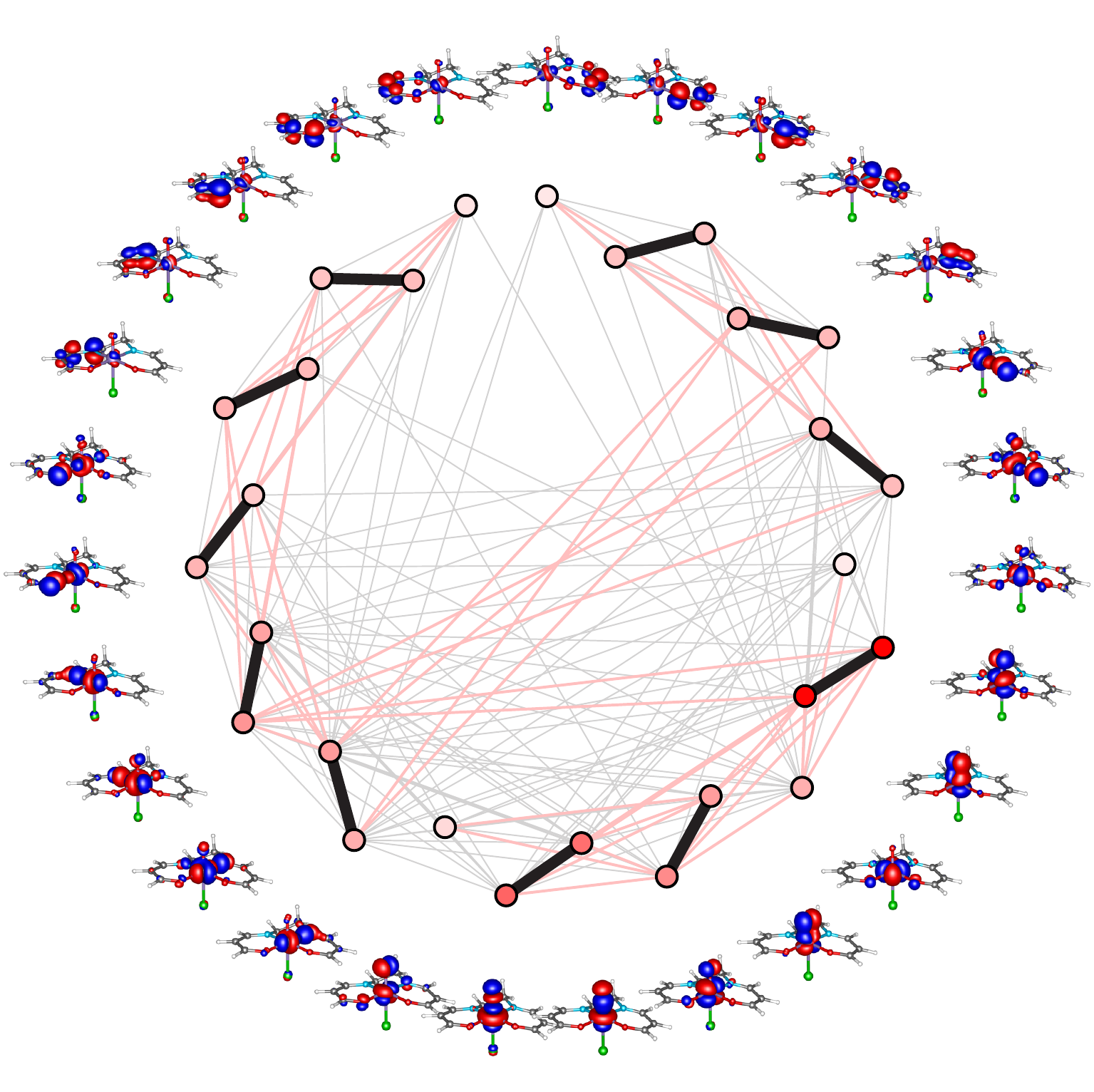} 
    \caption{Split-localized orbitals and their mutual information for $^1\text{A}$ (left) and $^3\text{A}$ (right)
    states of oxo-Mn(Salen) with the cc-pVDZ basis in CAS(28,27).
    The mutual information is color-coded: the black lines correspond to the strongest correlations ($10^{-1}$),
    pink ($10^{-2}$), and gray ($10^{-3}$).
    One-site entropy values are represented by a color gradient of the respective dot, red being the largest value
    and white being zero.}
    \label{fig_oxo_I}
  \end{center}
\end{figure*}

Based on the diagrams of mutual information plotted in Figure \ref{fig_oxo_I}, we assume that the orbitals
responsible for considerable decrease in energy are mainly the antibonding $\sigma^*$ and partially the 4d$_{yz}$ orbitals.
As expected, these are strongly correlated to the respective bonding orbitals.
Since their effect is the same for both states of interest, the remaining 4d$_{xy}$ has to cause
the change in the ground state.
This change can be then easily explained when we consider that in singlet state the valence
3d$_{xy}$ orbital is fully occupied and therefore the 4d$_{xy}$ orbital stabilizes the singlet via the
double-shell effect.
This claim is supported by the fact, that mutual information between the two orbitals is fairly
large in case of the singlet state.
Even though its value in the diagram is of order of magnitude $10^{-2}$, out of the remaining values of the same order
it is in fact one of the largest, namely 0.086, while in case of the triplet state its value is merely a fraction
i.e. 0.026.
Moreover, this orbital was also present in the study of Stein and Reiher \cite{stein_2016}, who obtained
the singlet ground state as well, with very similar energy difference between the two states.

\begin{table*}[!ht]
  \caption{The singlet and triplet energies of oxo-Mn(Salen) $E+2251$ in atomic units and the difference
  $\Delta E_\text{T-S}=E(^3\text{A})-E(^1\text{A}) $ in kcal/mol. Results for different active spaces
  and in various basis sets.}
  \def\arraystretch{1.2}
  \setlength{\tabcolsep}{0.5em}
  \begin{tabular}{lccrccrccr}
                     &  \multicolumn{3}{c}{cc-pVDZ} &  \multicolumn{3}{c}{cc-pVTZ} &  \multicolumn{3}{c}{cc-pVQZ} \\
                     & $^1\text{A}$ & $^3\text{A}$ & \mc{$\Delta E_\text{T-S}$}
                     & $^1\text{A}$ & $^3\text{A}$ & \mc{$\Delta E_\text{T-S}$}
                     & $^1\text{A}$ & $^3\text{A}$ & \mc{$\Delta E_\text{T-S}$} \\ \hline
DMRG-CASSCF(28,22)\cite{wouters_2014}
                     & $-0.7509$ & $-0.7593$ & $-5.3$\ph{0} &           &           &              &           &           &              \\
DMRG-CASSCF(26,21)\cite{stein_2016}
                     & $-0.7963$ & $-0.7954$ &  $0.6$\ph{0} &           &           &              &           &           &              \\
DMRG-CASSCF(28,22)\cite{sharma_2017}
                     & $-0.7991$ & $-0.8002$ & $-0.7$\ph{0} & $-0.9957$ & $-0.9926$ &  $1.9$\ph{0} & $-1.0449$ & $-1.0418$ &  $1.9$\ph{0} \\
DMRG-CASSCF(28,22)   & $-0.7502$ & $-0.7582$ & $-5.0$\ph{0} & $-0.9402$ & $-0.9473$ & $-4.5$\ph{0} & $-0.9891$ & $-0.9961$ & $-4.4$\ph{0} \\
DMRG-CASSCF(28,27)   & $-0.8803$ & $-0.8791$ &  $0.8$\ph{0} & $-1.0717$ & $-1.0697$ &  $1.2$\ph{0} & $-1.1210$ & $-1.1187$ &  $1.5$\ph{0} \\ \hline
TCCSD(34,25)\cite{veis_2016} \footnote{These values were obtained with split-localized ROHF orbitals.}
                     & $-2.7273$ & $-2.7330$ & $-3.6$\ph{0} &           &           &              &           &           &              \\
MRLCC(28,22)\cite{sharma_2017}
                     & $-3.2830$ & $-3.2600$ & $14.4$\ph{0} & $-4.1303$ & $-4.1310$ & $-0.5$\ph{0} &           &           &              \\
NEVPT2(28,22)\cite{sharma_2017}
                     & $-3.0109$ & $-2.9990$ &  $7.4$\ph{0} & $-3.8437$ & $-3.8463$ & $-1.6$\ph{0} & $-4.1441$ & $-4.1481$ & $-2.4$\ph{0} \\
LPNO-TCCSD(28,22)    & $-3.1455$ & $-3.1554$ & $-6.2$\ph{0} & $-3.9698$ & $-3.9798$ & $-6.3$\ph{0} & $-4.2479$ & $-4.2578$ & $-6.3$\ph{0} \\
LPNO-TCCSD(28,27)    & $-3.1491$ & $-3.1550$ & $-3.7$\ph{0} & $-3.9749$ & $-3.9798$ & $-3.1$\ph{0} & $-4.2531$ & $-4.2578$ & $-2.9$\ph{0}
\end{tabular}
  \label{tab_oxo}
\end{table*}

% EFFECT OF DYNAMIC CORRELATION
When dynamic correlation is added to the calculation by the LPNO-TCCSD method, a significant shift in
energies occurs.
Looking at the energies of the triplet state we can observe that the extra orbitals in the larger CAS seems to
contribute solely to the dynamic correlation at the CASSCF level, since the LPNO-TCCSD energies are virtually
the same for both active spaces.
Singlet energies, on the other hand, differ depending on the active space, which can be attributed to
the effect of aforementioned static correlation from the inclusion of 4d$_{xy}$ orbital.
Nonetheless, irrespective of the active space, we found out the ground state to be a triplet with slightly
larger singlet-triplet gap for the smaller CAS.
This is in agreement with our previously published results \cite{veis_2016}, in which we used canonical TCCSD,
but only with orbitals coming from an ROHF calculation.

% EFFECT OF THE LARGER BASIS SETS
Furthermore, these results remain consistent with larger basis sets.
Going from DZ to QZ basis at CASSCF level, the ground state of the particular CAS slightly stabilizes.
Although no significant change occurs for the singlet-triplet gap by including dynamic correlation in
the smaller CAS, the size of the gap decreases for larger CAS.
This change roughly corresponds to what was observed at CASSCF level.

Comparing these results to those published by Sharma et al. \cite{sharma_2017}, we conclude that employing
larger basis set does not qualitatively affect the ground state.
Since we have shown that the system is quite sensitive to composition of the active space and apart from DZ,
their energies agree very reasonably with ours, we suppose there might have been some inconsistency in the
active space for DMRG-CASSCF in the smallest studied basis set.
Especially our best result LPNO-TCCSD(28,27) is in excellent agreement with the best available NEVPT2(28,22)
in cc-pVQZ basis.

Regarding the future endeavors with this system, we would suggest the geometry reoptimization.
Although the used geometry serves as a useful benchmark for the reason that it is frequently used
by several different groups, since the introduction of dynamic correlation, the optimization
at CASSCF(10,10)/6-31G* appears to be fairly inadequate.

\section{Conclusions}
\label{sec_conclusions}

We introduced a new version of DMRG-TCCSD method, which employs the local pair natural orbital approach.
The method has been implemented in ORCA presently at the singles and doubles level.

We performed accuracy assessment of the method employing two systems, which were previously studied by
the canonical TCCSD method.
Regarding tetramethyleneethane, we were able to retrieve over 99.7\%  for triplet and over 99.8\%
for singlet state, while using the default settings of cut-off parameters.
For oxo-Mn(Salen), the amount of retrieved correlation was dependent on the size
of the active space used, ranging from 99.6\% for the larger CAS(28,27) to 99.8\% for smaller CAS(28,22).
Despite this dependence, an excellent agreement was achieved concerning the accuracy between the spin states.
Using the default settings resulted in singlet-triplet gap being off by 0.6 kcal/mol and with tighter cut-offs
only 0.04 kcal/mol compared to canonical calculation.

Furthermore, we investigated previously unexplored problem of varying reports of different ground states
of oxo-Mn(Salen) at CASSCF level.
We discovered that this inconsistency most likely originates from the composition of CAS.
In particular, we found out that the orbital responsible for stabilizing the singlet state is
the 4d$_{xy}$ 
orbital via the double-shell effect.
However, by employing the dynamic correlation treatment with LPNO-TCCSD, regardless of the basis set
the ground state was unambiguously found to be a triplet.

Regarding the future of the method, we would like to implement the DLPNO version of the TCCSD, which we hope
to further enhance capabilities of the method and also include
the perturbative triples correction for even more accurate results.

\section*{Acknowledgment}

We would like to thank Prof. Frank Neese for providing us with access to ORCA source code, as well as for helpful discussions,
Dr. Frank Wennmohs for technical assistance with the ORCA code, and Dr. Ond\v{r}ej Demel for helpful discussions.

This work has been supported by
the Czech Science Foundation (Grants No. 16-12052S and 18-18940Y),
Czech Ministry of Education, Youth and Sports (Project No. LTAUSA17033),
Charles University (Project GA UK No. 376217)
and
by The Ministry of Education, Youth and Sports from the Large Infrastructures for Research,
Experimental Development and Innovations project „IT4Innovations National Supercomputing Center – LM2015070“.

\bibliography{references,dmrg_tcc,cc,ors,dmrg,pno}
\bibliographystyle{achemso}

\end{document}